\documentclass[twocolumn,aps]{revtex4-1}
\usepackage{graphicx}
\usepackage{float}

\begin{document}

\title{Thermal conductivity reduction in rough silicon nanomembranes}

\author{G.Pennelli}
\email{g.pennelli@iet.unipi.it, tel. +39 050 2217 699, fax. +39 050 2217522}

\author{E.Dimaggio}

\author{M.Macucci}
\affiliation{Dipartimento di Ingegneria della Informazione, Universit\`a di 
Pisa, Via Girolamo Caruso 16, I-56122 Pisa, Italy}

\begin{abstract}
Nanostructured silicon is a promising material for thermoelectric conversion, 
because the thermal conductivity in silicon nanostructures can be strongly 
reduced with respect to that of bulk materials. We present thermal conductivity 
measurements, performed with the 3$\omega$ technique, of suspended 
monocrystalline silicon thin films (nanomembranes 
or nanoribbons) with smooth and rough surfaces. 
We find evidence for a significant effect of surface roughness on
phonon propagation: the measured thermal conductivity for the rough 
structures is well below that predicted by theoretical models which 
take into account diffusive scattering on the nanostructure walls. 
Conversely, the electrical conductivity appears to be substantially
unaffected by surface roughness: the measured resistance of smooth and rough 
nanostructures are comparable, if we take into account the geometrical factors.
Nanomembranes are more easily integrable in large area devices with respect
to nanowires and are mechanically stronger and able to handle much larger 
electrical currents (thus enabling the fabrication of thermoelectric devices
that can supply higher power levels with respect to current existing solutions).

\end{abstract}

\maketitle
\section{Introduction}
\begin{figure*}
\includegraphics[width=12 cm,keepaspectratio]{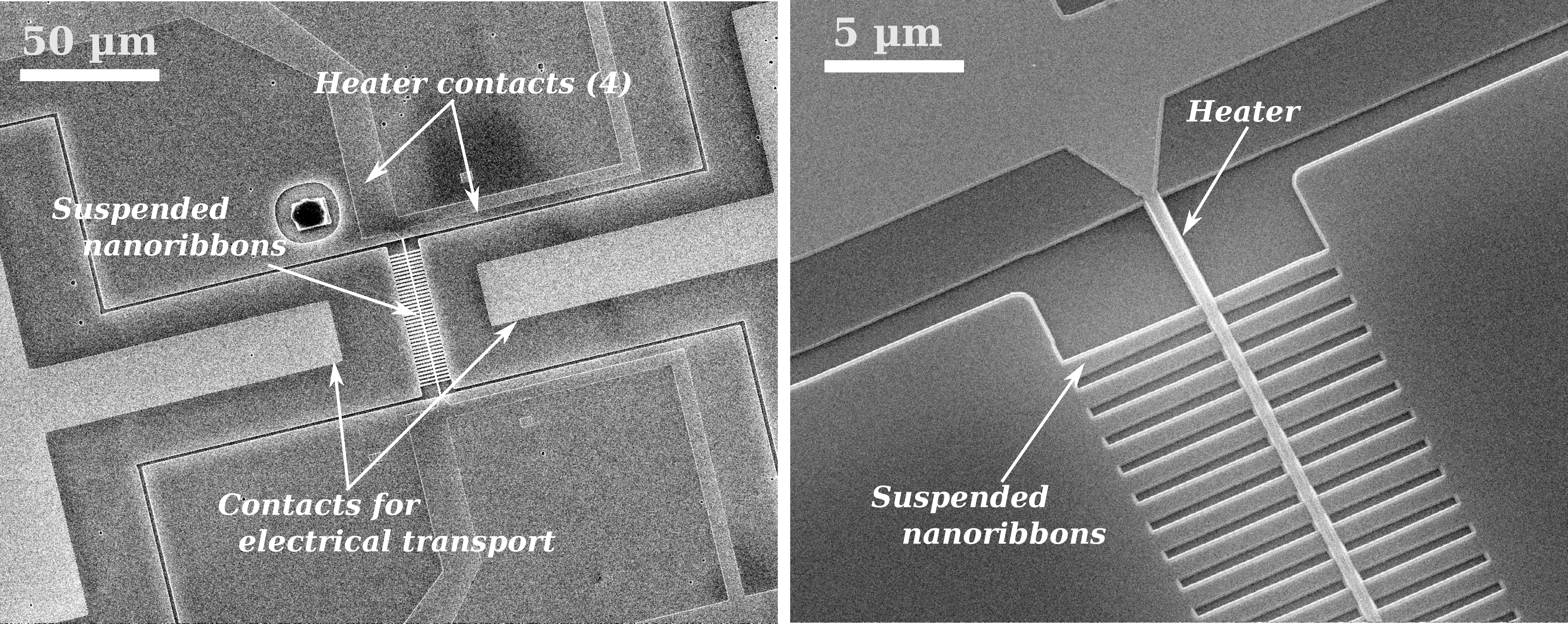}
\caption{SEM images of the suspended nanomembranes, used for the thermal and 
the electrical transport investigations.}
\label{SEM-photos}
\end{figure*}
In the last years, significant research effort has been devoted to the 
investigation of thermal transport in nanostructures, such as silicon 
nanowires\cite{li, hochbaum,  mio-conduzionetermica}. It has been found 
that their thermal conductivity $k_t$ is often rather small, because phonon
propagation is suppressed by scattering on the nanowire surfaces. As 
demonstrated by several experimental investigations\cite{lim-2012, park-2011, 
feser-2012,kim-2011}, and confirmed by
theoretical considerations\cite{martin-2009, liu, carrete-2011}, rough 
surfaces do suppress phonon conduction: values of thermal conductivity 
down by two orders of magnitude with respect to that of bulk silicon can 
in principle be achieved. 
The key point is that, in order to achieve satisfactory operation of 
thermoelectric devices, the electrical conductivity must instead be minimally
affected by surface roughness. Suppression of thermal conductivity while
keeping a good electrical conductivity opens up interesting opportunities in 
the field of energy harvesting, because a high thermoelectric conversion 
efficiency could be achieved with a material, 
such as silicon,  abundant on the Earth's crust and biocompatible. 
Since single nanostructures can handle only a limited amount of current, 
a thermoelectric generator 
(TG) useful for practical applications must consist in a large collection 
of interconnected nanostructures\cite{mio-nanonet-nanolet, mio-nanonet-eeng, fonseca-2012}. 
Silicon nanomembranes and nanoribbons, with one dimension $t_h$ (thickness) 
much smaller than the other (width $W$) represent an interesting alternative to
nanowires, since a very large number of them can be packed 
in parallel\cite{mio-generatore-orizzontale} with the larger side $W$ 
perpendicular to the surface. 
Although the nanomembrane approach is promising for practical TGs, it is 
expected that the thermal conductivity suppression with decreasing nanomembrane
thickness $t_h$ could be less effective than in nanowires\cite{marconnet-2013}.
Our study is aimed at understanding how far surface roughness can go in 
reducing thermal conductivity of nanomembranes. 

We developed a process for the fabrication of suspended silicon 
nanoribbons, together with contacts for the in-plane
electrical characterization and a heater for 3$\omega$ in-plane thermal transport 
characterization. We measured the
electrical and thermal transport of smooth nanoribbons, and compared them 
with that of rough ones. 
Roughness was characterized by means of AFM imaging. 
Thermal conductivity measurements are compared with simple 
theoretical models, based on the  assumption of fully diffusive phonon 
scattering on the nanomembrane walls. 
A strong reduction of the thermal conductivity in thin nanomembranes, with 
respect to that predicted by diffusive models, is demonstrated. Specific models 
for the phonon propagation in rough nanomembranes, which demonstrates these 
experimental results, still need to be developed.
 
\section{Device fabrication}
The left panel of Fig.~\ref{SEM-photos} shows a SEM image of a typical device,
made up of an array of suspended thin silicon membranes, arranged in a double 
comb configuration with a central 
silicon body. On this silicon body, a metal (gold) strip is the 
heater for the 3$\omega$ thermal conductivity measurement. The resistance 
of this metal strip can be measured by means of leads arranged
in a four contact configuration.  Another couple of contacts allow the 
measurement of the electrical conductivity 
of the silicon nanomembranes.  
The device has been fabricated on the top silicon layer of a Silicon On 
Insulator substrate. The top silicon layer is 
260~nm thick, the buried oxide layer is 1 $\mu$m thick and the total substrate 
thickness is 0.5~mm. The process is 
similar to that already used for the fabrication of silicon nanowire 
devices\cite{mio-nanofilo-2005, mio-set-2006, mio-pellegrini}, and is based 
on electron beam 
lithography\cite{mio-ebeam-2003} and silicon anisotropic etching. An eventual 
thinning of the top silicon layer is performed by dry oxidation and BHF etching. 
The thicknesses have been measured from Atomic Force 
Microscopy (AFM) images, taken during the process steps.
The process steps can be summarized as follows.\\ 
1) A 40 nm thick SiO$_2$ layer 
is grown on the top silicon layer. \\
2) Electron beam lithography is used for the definition of trenches in the 
top SiO$_2$ layer,
with the deposition of PMMA resist, e-beam exposure, development, 
and SiO$_2$ etching by means of Buffered HF (BHF). 
In this step, 
the comb geometry is defined in the top SiO$_2$ layer.  After BHF etching, 
the thickness of the top SiO$_2$ layer has been measured, with a result around 40 nm
for all the fabricated samples.
\begin{figure*}
\includegraphics[width=12 cm,keepaspectratio]{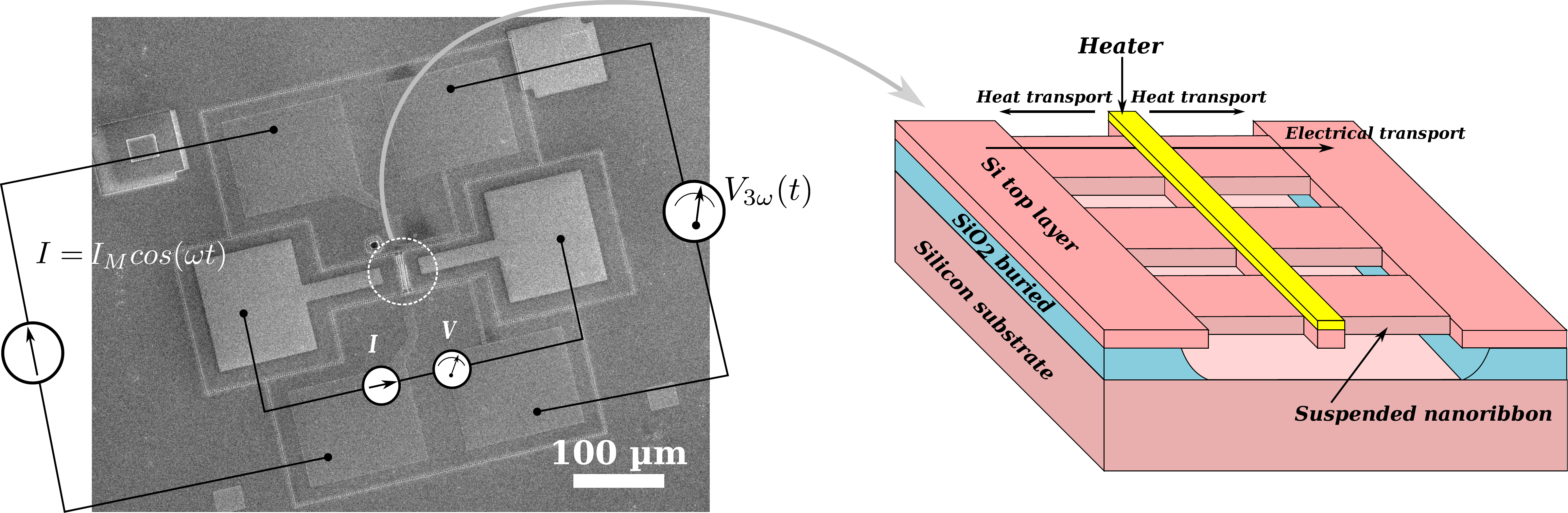}
\caption{Left panel: SEM image of the device. The sketches drawn on the image 
illustrate the measurement procedure.
The two contacts at the sides of the comb are used for the measurement of the 
electrical resistance of the
nanoribbons.The other four contacts are instead used for the precise 
measurement of the resistance of the metal
strip fabricated at the middle of the comb. Right panel: sketch 
of the device, showing the suspended
nanomembranes and the central part with the fabricated metal strip (heater). 
Thermal transport and electrical transport
can be investigated in the plane of the silicon nanomembranes.}
\label{SEM-sketch}
\end{figure*}\\
3) KOH etching has then been used for transferring this pattern to the 
top silicon layer. After this step, AFM imaging 
has been performed for the measurement of the thickness of the top silicon 
plus the top SiO$_2$ layer. Therefore, the
top silicon layer thickness has been obtained by subtracting the 
SiO$_2$ thickness, measured before 
the KOH etching, as KOH has practically no effect on SiO$_2$. A final 
nanomembrane thickness of 240 nm has been measured, if the process
is performed on the substrate with a top silicon layer 260 nm thick. Different
thicknesses (see the measurement results) have been achieved by thinning the
substrate through oxidation/BHF etch before the first lithographic step.\\
4) Then the top silicon dioxide layer has been removed by means of BHF 
etching. This etch step allows the 
deposition of the metal tracks in direct contact with silicon, without any 
insulating layer in between. \\
5) An e-beam lithography step has then been performed on a PMMA double layer, 
and then a metal layer, consisting of 10 nm of Chromium 
for adhesion and of of 80 nm of Gold, has been deposited by means of 
thermal evaporation. Metal lift off has been performed in hot acetone. 
This lithographic step has
been carefully aligned with the 
comb structure underneath, so that the metal track has been positioned exactly 
on the silicon body in the middle of the device. \\
6) The nanomembranes (both the comb and the central body) have been suspended 
by underetching the buried oxide by means of BHF. 
The final device is shown in Fig.~\ref{SEM-sketch}, together with a sketch 
of the suspended structures. \\
The sketch, drawn on the SEM photo, illustrates the purpose of each contact. 
The contacts at the sides of the comb 
are used for the measurement of the overall electrical conductivity of the 
nanoribbons. The four contacts are used for 
the precise measurement of the resistance of the metal strip fabricated in 
the middle of the comb. 
The nanomembrane roughening is performed after the definition of the metal 
track, but before the suspension
of the nanomembranes by means of SiO$_2$ underetching: nanomembranes are 
very delicate, and any attempt at roughening suspended
nanostructures has resulted in a very low yield in terms of final number 
of working devices. 

\begin{figure}
\includegraphics[width=6 truecm,keepaspectratio]{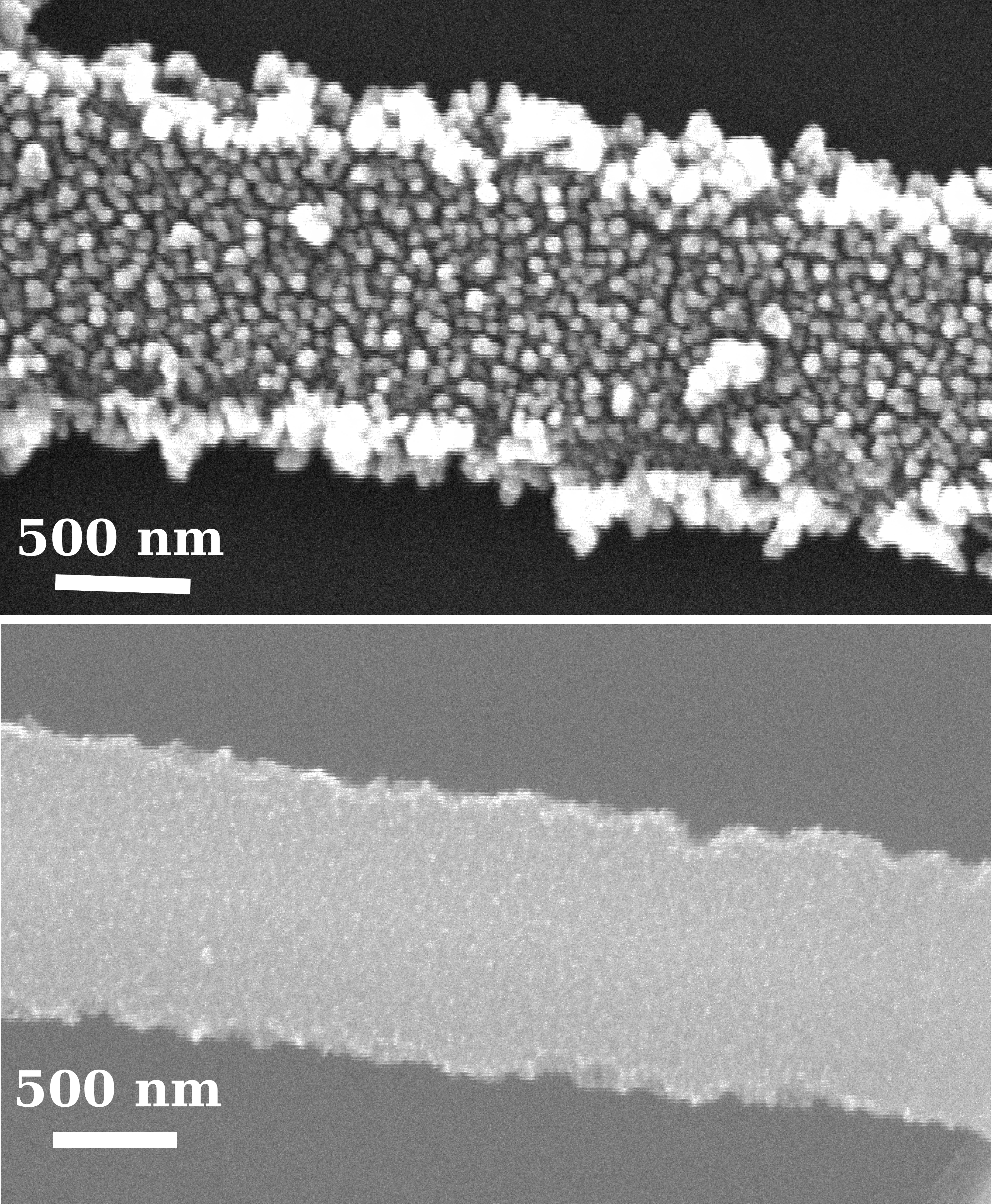}
\caption{Top image: SEM image of a silicon nanoribbon, covered by Silver 
nanoparticles after the MACE step. Bottom
image: SEM image of a silicon nanoribbon, once the Silver 
nanoparticles have been removed by means of HNO$_3$ etching; the surface
appears to be very rough. }.
\label{SEM-nanoparticles}
\end{figure}
A rough surface has been achieved by soaking the sample in a solution with 
hydrofluoric acid and silver nitrate; this technique 
has been already used by Lim\cite{lim-2012} and coworkers for roughening 
silicon nanowires. The SEM image of 
Fig.~\ref{SEM-nanoparticles} shows silver nanoparticles, deposited on the 
surface of a nanomembrane by
soaking the sample in a solution AgNO$_3$/HF 0.01/5.1 mol for 60 s. Silver 
is reduced, withdrawing 
an electron from silicon (i.e. injecting a hole), so that it precipitates, 
aggregating in nanoparticles on the 
silicon surface. The reduced silicon is etched by the hydrofluoric acid. Therefore, 
a local etching of silicon occurs under the silver nanoparticles. This mechanism is 
not fully understood yet, even if it is the basis of the
Metal Assisted Chemical Etching (MaCE) technique, largely used for the 
fabrication of silicon nanowires 
perpendicular to a silicon substrate\cite{cu-deposition-nanoletters}.
We use this MaCE technique to produce shallow holes that are randomly 
distributed on the silicon surface. After the etch step, the silver 
nanoparticles 
have been removed by soaking the sample in HNO$_3$: H$_2$O 1:5 for 2 m.
The result is a very rough surface, as visible in the SEM image in the bottom
panel of Fig.~\ref{SEM-nanoparticles}. 
The $I-V$ characteristic of the silicon nanoribbons has been measured before 
and after the roughening process, through the
contacts placed at the sides of the comb, as shown in Fig.~\ref{SEM-sketch}. 
In this way, the relationship between the electrical conductivity before 
and after roughening 
has been established. An important quantity that must be determined for the 
evaluation both of the electrical and of the thermal conductivity is the 
final thickness of  the silicon nanomembranes. The roughening process leaves 
a chaotic pattern on the silicon surface, but it also removes a small amount 
of silicon, thus reducing the thichness of the nanoribbons.
As HF etches the SiO$_2$ buried 
layer, a direct measurement of the final
thickness of the nanomembranes, after the roughening process, is not possible. 
Therefore, the thickness has been measured using the gold metal tracks as a
reference, because gold is substantially unaffected by the etching processes.
The thickness of the metal tracks, with respect to the silicon, has been 
measured by means of AFM imaging before and after 
the roughening process. The difference between the two measurements 
corresponds to the amount of silicon that has been removed. Due to the 
roughness of the silicon surface, an average thickness has been obtained 
by averaging over areas of several square microns.

The thermal conductivity is evaluated with the 3$\omega$ technique, 
exploiting the metal structure fabricated in the middle of the double comb
both as a heater and as a temperature sensor. A sinusoidal 
electrical current is injected into the metal track through two of the four available 
contacts. The heat generated as a result of the Joule effect can be dissipated 
only through the nanomembranes in the direction of their plane, because they 
are suspended. 
The voltage is measured through the other two available contacts 
(four-probe configuration), and the amplitude
of the third harmonic is extracted by means of a lock-in amplifier 
(see the sketch on the SEM image of Fig.~\ref{SEM-sketch}). 
As it has been demonstrated in several applications, the amplitude of the
third harmonic is related to the thermal
dissipation\cite{cahill-1990, zhang-2001, goodson-2008}, which in our case 
is due to the thermal conductivity in the film plane.
For the proper application of the 3$\omega$ technique it is essential to choose
the correct model for data reduction. Although an analytical model
has been developed for 2D structures\cite{goodson-2008} (thin membranes), 
in our case we have fitted the data by means of an approach based on
FEM (Finite Element Method) simulations that we have 
developed~\cite{mio-3omega-misura}, using the 
thermal conductivity as a fitting parameter. 
The simulation takes into account the thermoelectric
transport in the metal and in the silicon nanostructures, considering 
the measured electrical conductivities.
\begin{figure*} 
\includegraphics[width=12 cm,keepaspectratio]{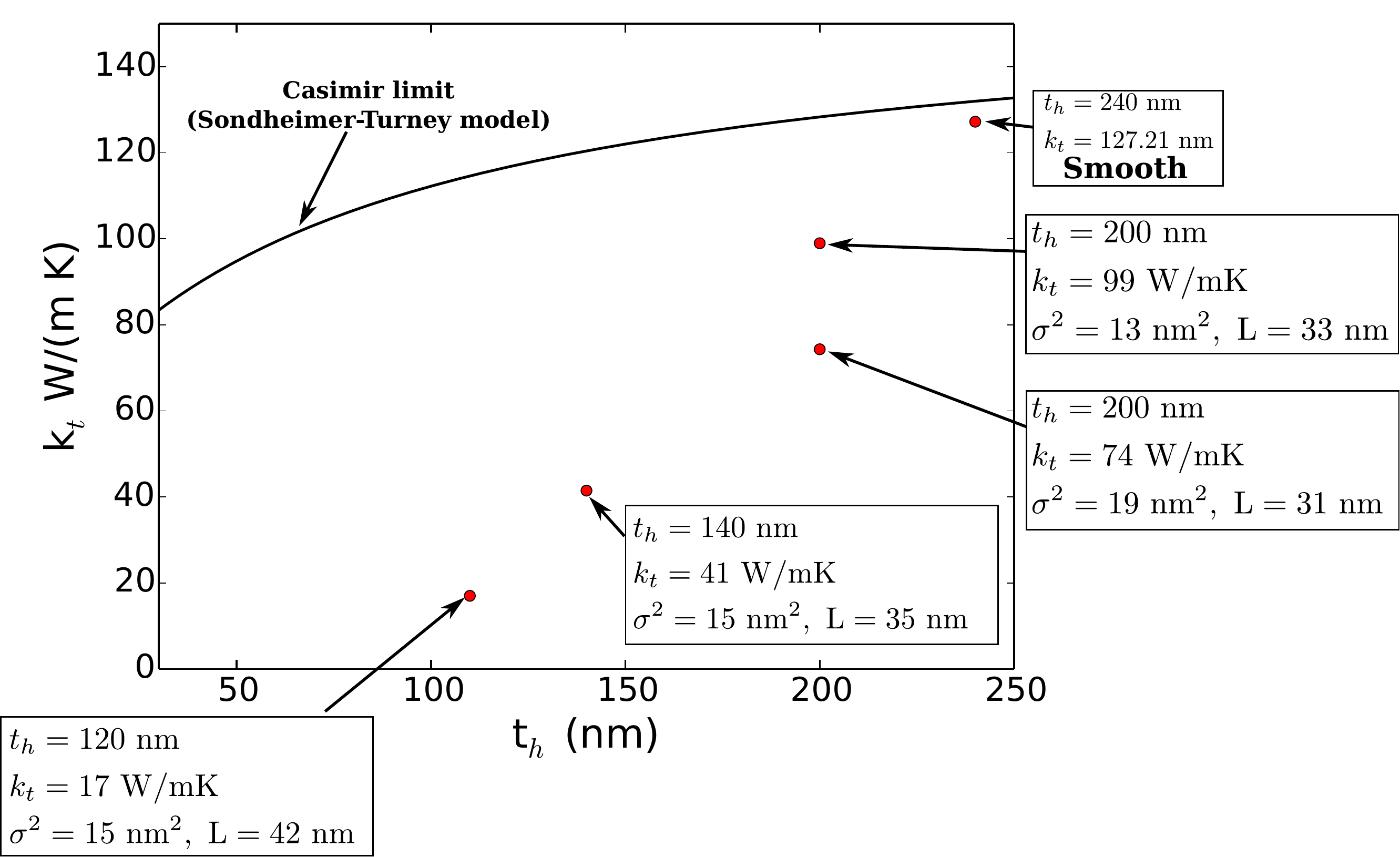}
\caption{Comparison between a theoretical model (see text) for thermal 
conductivity in nanomembranes and nanowires and our experimental results.}
\label{casimir-limit}
\end{figure*}

\section{Thermal and electrical conductivities}
In Fig.~\ref{casimir-limit} the results of the measured thermal conductivity
of smooth and rough silicon nanomembranes are shown. On the same plot,
the thermal conductivity predicted by means of a theoretical 
model is also reported for comparison purposes.
The theoretical evaluation of the thermal conductivity $k_t$ has been 
performed following the Callaway-Holland formalism\cite{callaway-1958, holland-1963}, 
in the vision developed by Asen-Palmer\cite{asen-1997} and, in particular, from 
the expression by Morelli {\sl et al.}\cite{morelli-2002}: 
\begin{equation}
k_{t}=\frac{1}{3}\frac{1}{2\pi^{2}}\frac{k^{4}T^{3}}{\hbar^{3}v_{s}}
\int_{0}^{\frac{T_{\theta}}{T}}\tau(x)\,\frac{x^{4}e^{x}}
{\left(e^{x}-1\right)^{2}}\mathrm{\mathrm{d}}x
\end{equation}
where $T_{\theta}$  is the Debye temperature, $\hbar$ is the reduced Planck 
constant, $T$ is the 
absolute temperature, $k$  is the Boltzmann constant, $x=\hbar \omega /kT$, 
and $\tau(x)$ 
is the relaxation time. 
Both the contribution of $k_t^L$ 
due to the longitudinal mode and that of $k_t^T$ due to the transverse 
phonon modes have been evaluated taking into 
account the proper parameters, for example considering $T_\theta^L=586$~K and 
$T_\theta^T=240$~K for the 
Debye temperatures of the longitudinal and transverse phonons, respectively. 
The two 
contributions have then been combined to derive the total thermal 
conductivity:
$k_t=k_t^l+k_t^t$.

\begin{figure*}
\includegraphics[width=12 cm,keepaspectratio]{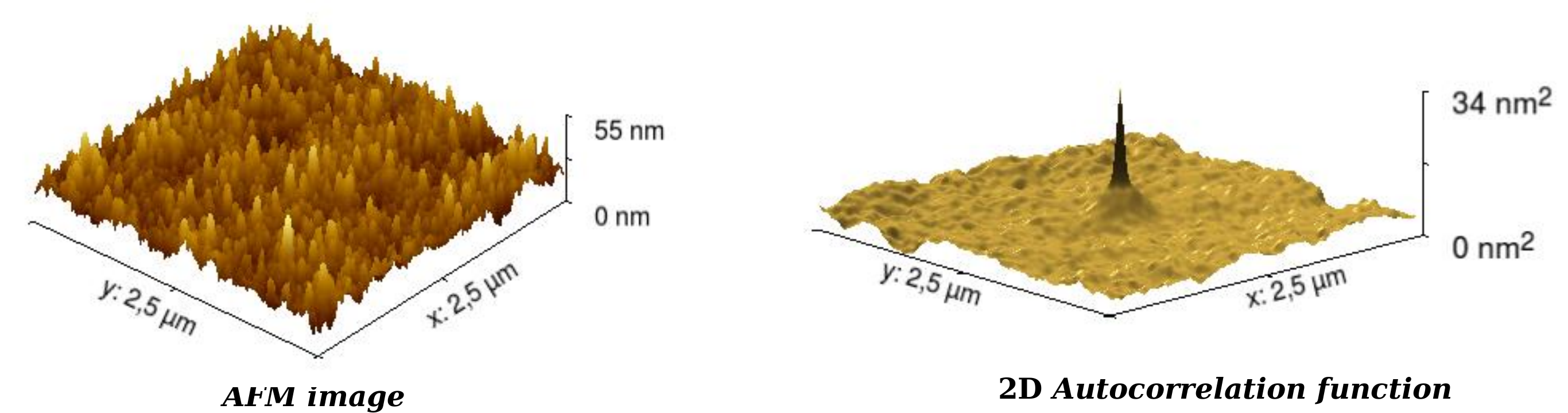}
\caption{Left panel: AFM image of the surface after the roughening process. 
Right panel:
two-dimensional autocorrelation function of the surface roughness.}
\label{AFM-image}
\end{figure*}
In the evaluation of the relaxation time $\tau(\omega)$, we have taken 
into account the contribution
of different scattering processes, combining them by means of 
Matiessen's rule. 
 
The relaxation times $\tau_N(\omega)$, relative to the phonon-phonon normal
scattering, $\tau_U(\omega)$, relative to the phonon umklapp scattering 
and $\tau_I(\omega)$, relative
to the isotope scattering, have been evaluated using the parameters reported 
by Morelli {\sl et al.}~\cite{morelli-2002} for the longitudinal and the 
transverse phonon modes. 
Several models have been developed in the literature for the inclusion of 
the phonon boundary 
scattering, which is a phenomenon of fundamental importance for the 
accurate evaluation of the thermal conductivity 
in nanostructures. In our model we have considered completely diffusive 
boundary scattering, so that the resulting thermal conductivity is the one
corresponding to the so-called Casimir limit.   
In order to include also the contribution from boundary scattering, 
we followed an approach similar to the one proposed by 
Sondheimer\cite{sondheimer-1952} for
the electrical conduction
in thin metal films, based on the solution of the
Boltzmann transport equation. 
For the present work, we have used the simpler solution
developed by Turney {\sl et al.}\cite{turney-2010}
for phonon transport in thin silicon layers with a width much greater than
the thickness $t_h$.
Following this model, at first a preliminary relaxation time
$\tau^{'}(\omega)$ has been computed with Mathiessen's rule, including
all the scattering phenomena, but the phonon boundary scattering. Then,
the total relaxation time has been evaluated as (see the paper by Turney {\sl
et al.}\cite{turney-2010}, with the probability of
thermalised boundary scattering $p=1$):
\begin{equation}
\tau(\omega)=\tau^{'}(\omega)\left(1-{1 \over \delta}\right)\left(1-e^{-\delta}\right)
\end{equation}
where $\delta=t_h/v_s \tau^{'}(\omega)$ ($v_s$ is $v_s^L$ for longitudinal
and $v_s^T$ for transverse phonon).

In Fig.~\ref{casimir-limit} the continuous line represents the thus 
computed thermal conductivity $k_t$ as a function of the thickness $t_h$, 
for a silicon nanoribbon 1 $\mu$m wide, which is a width comparable to that 
of our devices. It is possible to see that with the adopted model $k_t$ 
exhibits a moderate dependence on the thickness of the nanoribbon,
because the mean free path $\lambda$ for boundary scattering is mainly  
determined by the nanoribbon width rather than the thickness, as already 
reported by Wang and Mingo\cite{wang-mingo-2011}. 
%
The experimental value reported at the top right of Fig.~\ref{casimir-limit}
has been obtained on a 240~nm thick smooth nanoribbon, and it appears to be in
good agreement with the theoretical prediction.
Indeed at the end of the 
fabrication process the nanoribbons are very smooth if the roughening 
step has not been performed. 
The oxidation step used for the definition of the mask for silicon etching 
preserves the smoothness of the surface, while reducing the thickness of the 
nanoribbons, whose final value is 240~nm (from AFM measurements). 

The thermal conductivity undergoes a dramatic change if the roughening 
process (described in the previous Section) is performed.
The AFM image in the left panel of  Figure~\ref{AFM-image} shows the 
surface after the roughening process. In the right panel we report the 
two-dimensional autocorrelation function 
of the same image, from which the main characteristics of the roughness 
can be derived. 
In particular, the autocorrelation function has been fitted with a gaussian: 
$g(\overline r) = \sigma^2~e^{-{r^2 \over L^2}}$, where $\sigma$ is the 
standard deviation
with respect to the average value of the surface and $L$ is the autocorrelation parameter. For the
sample shown in Fig.~\ref{AFM-image} we achieved $\sigma^2=17$ nm$^2$ 
and $L=42$ nm.
Part of the silicon is etched during the roughening process, therefore at 
the end the final
thickness of the nanomembranes is reduced: the final thickness is measured
by means of AFM imaging, as explained in the previous section.

For a final thickness of 200~nm we report the values of the thermal 
conductivity for two different values of the roughness variance: while for 
$\sigma^2=13$~nm$^2$ we get 99~W/(m K), for $\sigma^2=19$~nm$^2$ $k_t$ is 
reduced down to 74~W/(m K). As the thickness is decreased we observe, for
similar values of the roughness, a sharp decrease of the thermal conductivity,
down to 17~W/(m K) for $t_h=120$~nm. 
This is analogous to what has been 
measured\cite{li, hochbaum,  mio-conduzionetermica,lim-2012, park-2011, feser-2012, kim-2011} and theoretically 
predicted\cite{martin-2009, liu, carrete-2011} for silicon nanowires.

It must be noted that the roughening step must be performed before 
the underetch of the buried oxide, which is needed to achieve the 
suspension of the nanomembranes. 
Therefore, we cannot measure the thermal conductivity before and after 
the roughening process, and the values reported in Fig.~\ref{casimir-limit} 
have been obtained on different samples.\\
The electrical resistance
has been measured through the electrical contacts fabricated
at the sides of the suspended nanoribbons (see Fig.~\ref{SEM-photos}).
The top silicon layer has an uniform doping of $5\times10^{18}$ cm$^{-3}$. 
We report, as a typical example, the resistance of the sample with smooth 
nanoribbons 240 nm thick: the resistance was $R=246.49$ $\Omega$. For the 
rough sample, with nanoribbons 120 nm thick, we measured an the electrical 
resistance of $R=110.35$ $\Omega$: this is compatible with the final 
thickness of the sample (120 nm), and with the slight reduction of the 
nanoribbon width as a consequence of silicon etching associated with roughening. 
Therefore, the final measured resistance is consistent with the variation 
of the geometrical parameters (thickness and width), and this
means that the electrical conductivity is practically unchanged. Both
resistance values, taking into account the geometrical factors, give an 
electrical conductivity of 0.006 $\Omega^{-1}cm^{-1}$. This value is
compatible with the nanoribbon doping.

This is a key point for future applications of the monocrystalline nanoribbons 
for thermal to electrical energy conversion: the thermal conductivity
is strongly reduced by the rough surface, while the electrical conductivity 
is practically unaffected.

\section{conclusions}
We have presented an application of the particular version of the 
3$\omega$ technique that we have developed for the evaluation of the thermal 
conductivity of rough suspended silicon nanomembranes, for which a direct
application of the commonly used analytical approximations would lead to 
excessive approximations.
While in the absence of significant surface roughness the results of our
measurements are in agreement with theoretical predictions, the introduction
of roughness appears to strongly suppress thermal conductivity, in analogy with
what has been previously observed in the literature for silicon nanowires, while
substantially leaving the electrical conductivity unaffected. If confirmed,
this would make rough silicon nanoribbons quite interesting candidates for 
silicon-based thermoelectric generators, because of the relatively high 
figure of merit $ZT$ and the current carrying capability, which is significantly
larger than that of silicon nanowires.

\bibliography{tre-omega-dati}

\end{document}